# Structural and electrical properties of electrodeposited single junction of cuprous (I) oxide – copper

Edward Rówiński[1,a], Mateusz Pławecki[1,b]

[1]*Institute of Materials Science, University of Silesia, 75 Pułku Piechoty 1A, 41-500 Chorzów, Poland*

[a]edward.rowinski@us.edu.pl, [b]mplawecki@us.edu.pl

**Abstract.** Cuprous (I) oxide ($Cu_2O$)-based solar cells were fabricated with the use of the electrodeposition technique at nanometre-scale, and the structural, morphological and electrical properties were investigated. The $Cu_2O$ layers were electrodeposited on crystalline and polycrystalline copper substrates. To complete the $Cu_2O/Cu(100)$ and $Cu_2O/Cu$ interfaces as the solar cells the top electrodes of silver paste were painted on the rear of $Cu_2O$. The microscopic analysis exhibits an uneven surface morphologies of a $Cu_2O$ film with the roughness of 92.5 nm, while the X-ray diffraction analysis reveals that the layers are $Cu_2O$-type polycrystalline structures with the thickness of 493 nm and the crystallite size of 69.8(6) nm. The theoretical analysis of the current–voltage curve was provided to determine the values of electrical parameters of the most efficient solar cell of $Ag/Cu_2O/Cu(100)$ and clearly indicate presence of two Schottky barriers at interfaces.

## 1. Introduction

The solar cell materials have been the subject of intense research for many years due to their unusual optical and electrical properties [1-12]. On the basis of theoretical studies it may be proven that , photovoltaic technology has advanced considerably, resulting in single-junction solar cells with a record efficiency of 28.3% and multi-junction cells with an efficiency (under concentrated illumination) of 43.5% [2]. The initial effort to develop a solar energy conversion device can be traced back to the metal/semiconductor (MS) junction with a thin metal film [6,7]. The efficiency of $Cu/Cu_2O$ Schottky barrier solar cells has remained far below the theoretical value. The best results obtained so far remain in the range of 1-2 % [7].
Motivated by these results, the authors of presented studies aim to show how to explain crystalline and electrical behaviors of solar cell based on a nanocrystalline $Cu_2O$ layer sandwiched between two electrodes that consist of polycrystalline silver and crystalline or polycrystalline copper.

## 2. Materials and Methods

Two samples were constructed in sandwich geometry with each of them having a dimension of 1 x 1 cm$^2$. The $Cu_2O$ layers were prepared on crystalline Cu(100) (5 mm thickness) and polycrystalline Cu (0.5 mm thickness) substrates by electrodeposition using platinum counter electrode. Copper (II) sulfate (CuSO4, 0.4 mol L$^{-1}$, Wako 97.5 %) and L-lactic acid (3 mol/ L$^{-1}$, Wako) were dissolved into distilled water. Electrolyte pH was adjusted to 12.5 by adding NaOH. The electrolyte temperature was kept at 60 ºC during electrodeposition. The current density of electrodeposition was carried out at 1.5 mA/cm$^2$ and at the time of 10 min. Close to the value of current density, deposition time and Faraday's law, the estimated thickness of the $Cu_2O$ layers was about 0.5 nm.
To complete the interfaces the single solar cell a top electrode of silver paste (TAAB, S 270) was painted on the rear of $Cu_2O$. The $Cu_2O$ serves as a common active layer to two electrodes.
      The structural studies were carried out by X-ray diffraction (XRD) and grazing incidence X-ray diffraction (GIXD) on an Empyrean PANalytical powder diffractometer using Cu K$\alpha_1$ radiation. A grazing incidence geometry with an incident angle of 1 degrees was applied. The backscatter Laue diffraction pattern was provided on XRT-100 CCM diffractometer, company EFG. All measurements were taken at the room temperature.
The atomic force microscopy (AFM) image and micrograph were determined by a Hysitron Ti 950 Triboindenter equipped with a Q-Scope 250 and JEOL-JSM-6480 ($U$=20 kV) scanning electron microscope, respectively. All measurements were carried out at the room temperature.

Measurements of emission and excitation spectra were recorded using a FluoroMax-4 spectrofluorometer (Horiba, USA), equipped with automated polarizers. The 150 W ozone-free xenon arc lamp is source blazed at 330 nm (excitation) and 500 nm (emission).

The current –voltage curves were carried out using 2400 Series Source Meter, Keithley Instruments. The source was a 75 W wolfram lamp equipped with a sunlight filter to match the emission spectrum of the lamp to 1000 W/m$^2$. The solar cells were masked with a metal aperture to define an active area of 0.01 cm$^2$.

## 3. Theoretical model of current-voltage characteristic for metal-semiconductor nanostructure – metal solar cell

For the sake of simplicity, we propose a versatile theoretical approach based on photonics and collection losses due to incomplete light refraction and reflection. The current of metal-semiconductor-metal solar cell as a function of the voltage can be expressed as follows [1,4,6-12]:

$$I(U) = \sum_{i=1}^{2}[I_i \cdot (\mathrm{Exp}\{\frac{q \cdot [U - I(U) \cdot R_i^s]}{k_B \cdot T \cdot n_i}\} - 1) + \frac{U - I(U) \cdot R_i^s}{R_i^{sh}}] - I_{sc} \qquad (1)$$

$$I_i = A_i \cdot A_i^{**} \cdot T^2 Exp[-E_i^{SBH}/(k_B \cdot T)] \qquad (2)$$

$$I_{sc} = Q \cdot \{1 - [(1-R) \cdot L + R]\} \cdot [1 - \mathrm{Exp}(-\alpha \cdot d)] \cdot q \cdot n_{ph}(E_g) \qquad (3)$$

$$Q = Q_{cell} + Q_{col} \qquad (4)$$

$$Q_{col} = (1 + L/2)\frac{(1-R)^2}{m}; \qquad L = 1 - \frac{\sqrt{n^2 - 1}}{n}; \qquad R = \frac{n-1}{n+1} \qquad (5)$$

where $I(U)$ - the total current, $U$ - voltage, $I_{sc}$ - the short circuit current (photo-generated current), i - the number of interfaces, $I_i$ - the effective saturation current, $q$ - the elementary charge, $k_B$ -the Boltzmann constant, $T$ -the absolute temperature, $n_i$ - the ideality factor, $R_i^{sh}$ - the shunt resistance, $R_i^s$ - the series resistance, $A_i$ - the area of detector, $A_i^{**}$ - the Richardson's constant, $E_i^{SBH}$ - the Schottky barrier height, $Q$ - the total photovoltaic efficiency, $Q_{cell}$ - the collection efficiency defined as the ratio of the carriers passing through the circuit to those which have been generated in the layer, $Q_{col}$ - the efficiency of the collector through the two boundary planes, which is given by Eq.(5), $R$ - the reflection coefficient, $1-R$ - the fraction of light entering in the system, $(1-R) \cdot L$ - the loss due to the fluorescence process, $m$ - the number of parts containing an equal number of photons, $n$ - the refractive index of semiconductor, $\alpha$ - the absorption constant, $d$ - the thickness of the absorbing semiconductor layer, $E_g$ - the band gap energy of the absorbing semiconductor, $n_{ph}(E_g)$ - the number of photons per second per unit area of a single junction whose energy is great enough to generate hole-electron pairs in the semiconductor layer.

## 4. Results and discussion

The aim of the research was to gain insight into the nature of the crystalline behavior of the solar cells by means of grazing incidence X-ray diffraction and X-ray diffraction. It has been found that the Cu$_2$O layers are polycrystalline with no traces of CuO (Fig.1a). XRD lines correspond to Cu$_2$O phase and the polycrystalline Cu substrate (Fig.1a). The particle size was estimated by using the well-known Scherrer's equation [2,3]. The crystallite sizes of Cu and Cu$_2$O were determined to be 49.2(3) nm and 69.8(6) nm, respectively. The last value of crystallite size of Cu$_2$O also denotes the p-type layer conductivity [3,13]. Figure 1b shows the uneven surface morphology of Cu$_2$O layer obtained from a scanning electron microscope (SEM). Figure 1c shows the backscatter Laue diffraction pattern from the crystalline (100) copper. This indicates that the Laue diffraction pattern is very symmetric, with sets of diffraction spots arranged in rings revolving around the center of the pattern. The roughness of the active layer of Cu$_2$O deposited on the Cu(100) substrate was evaluated with the use of atomic force microscopy (AFM), and an image of a typical scan is provided in Fig. 1d, which demonstrates a film with the roughness of 92.5 nm.

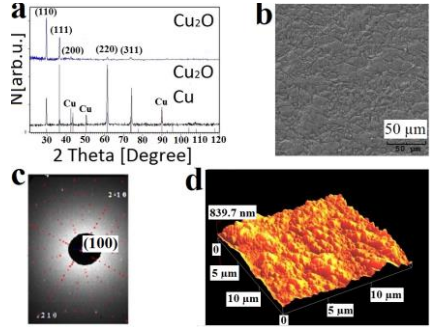

**Fig. 1.** Characteristics of $Cu_2O$ layer deposited on the polycrystalline and crystalline copper substrates. **a,** GIXD at $1^o$ incidence angle of the $Cu_2O$ layer and XRD patterns of the $Cu_2O$ layer and the polycrystalline copper substrate (N is the intensity expressed in arbitrary units, arb.u.). **b,** Micrograph of $Cu_2O$ layer obtained from a scanning electron microscope. **c,** The backscatter Laue diffraction pattern from the crystalline (100) copper substrates. **d,** AFM image of $Cu_2O$ deposited on the crystalline (100) copper substrate.

Figure 2a shows the photoemission spectrum of the $Cu_2O$/Cu(100) interface. The spectrum has two peaks situated at $E_a$~ 2.03 eV and $E_b$~ 2.17 eV, and the broad four peaks around $E_c$~ 2.65 eV, $E_d$~ 2.74 eV, $E_e$~3.17 eV and $E_f$ ~3.23 eV, respectively. The result is in accordance with the theoretical peaks in the density of states of $Cu_2O$ [13]. Figure 2b reveals that the blue-shift behavior may be then used as an indicator of the absorption of phonons by some electrons transferred to the highest level. The anti-Stokes shift is about 0.16 eV. Figure 2c shows the calculated current-voltage result obtained from this model and the experimental current-voltage characteristic for the Ag/$Cu_2O$-layer/Cu(100) solar cell. In this approach, the Schottky barrier heights of 1.1 eV for the Ag/$Cu_2O$ interface and 1.4 eV for the $Cu_2O$/Cu interface were reported, which are acceptable when compared to the Schottky-Mott model [1]. It is indicated here that the $U_{oc}$ depends linearly on the magnitude of the Schottky barrier height (Fig. 2d). As a result of the fact that $n_{Cu_2O/Cu} = 1$ the rectifying (homogenous) junction exists at the $Cu_2O$/Cu interface in the case of the back wall cells. In the case ($n_{Ag/Cu_2O} = 1.35$) of the front wall cells the inhomogeneous (rectifierlike) junction exists at the interface between the silver and $Cu_2O$ layers. Thus, theoretical result reproduces the experimental curve. The chosen values of solar cell parameters, listed in Table 1, were obtained through the current-voltage measurement.

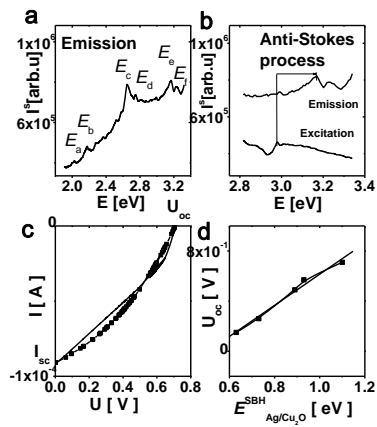

**Fig. 2.** Fluorescence emission spectrum and anti-Stokes emission process between the absorbed and emitted photons of the $Cu_2O$/Cu(100) interface. Current-voltage characteristics of the Ag/$Cu_2O$/Cu(100) solar cell, open-circuit voltage as a function of Schottky barrier height of Ag/$Cu_2O$ interface for the Ag/$Cu_2O$/Cu(100) solar cell. **a,** Fluorescence emission spectrum recorded for a typical $Cu_2O$/Cu interface. $I^s$ is the fluorescence intensity expressed in arbitrary units (arb.u.). The peaks, designated with the energies from '$E_a$' to '$E_f$', are due to emission of photons. **b,** Fluorescence emission and excitation spectra of the $Cu_2O$/Cu interface. Energy difference between the positions of the band maxima of emission and excitation spectra is represented by the blue shift of ~0.16 eV due to anti-Stokes process. **c,** Experimental (dot line) and theoretical (solid line) current-voltage characteristic of the Ag/Cu$_2$O/Cu(100) solar cell using Eqs. (1)-(5). The parametric values of the model were found to be i = 1 (=Ag/Cu$_2$O), i = 2 (=Cu$_2$O/Cu), $A_{Ag/Cu_2O} = 10^{-6}$ m$^2$, $A^{**}_{Ag/Cu_2O} = 1.2 \cdot 10^6$ Am$^{-2}$K$^{-2}$, $E^{SBH}_{Ag/Cu_2O} = 1.1$ eV, $k_B T = 0.026$ eV, $A^{**}_{Cu_2O/Cu} = 0.8 \cdot 10^6$, $E^{SBH}_{Cu_2O/Cu} = 1.4$ eV, $E_g = 2.1$ eV, $n_{ph}(E_g) = 5 \cdot 10^{21}$ s$^{-1}$m$^{-2}$, $n = 2.54$, $R = 0.43$, $L = 0,08$, $m = 5$, $Q_{col} = 0.06$, $Q_{cell} = 0.22$, $Q = 0.28$, $\alpha = 13$ cm$^{-1}$, $n_{Ag/Cu_2O} = 1.35$, $n_{Cu_2O/Cu} = 1$, $R^{sh}_{Ag/Cu_2O} = 10$ k$\Omega$, $R^{sh}_{Cu_2O/Cu} = 44$ k$\Omega$, $R^{s}_{Ag/Cu_2O} = 50$ $\Omega$, and $R^{s}_{Cu_2O/Cu} = 100$ $\Omega$. **d,** Open-circuit voltage as a linear function of Schottky barrier height of Ag/Cu$_2$O interface for Ag/Cu$_2$O/Cu solar cell. The coefficients of determination for the fitted data are much larger than 0.98.

**Table 1** Solar cell parameters of two devices based on the same thicknesses of Cu$_2$O layers. Relative errors were less than 2%.

| Solar cell device (sample) | Open-circuit voltage ($U_{oc}$) [V] | Short-circuit current density ($J_{sc}$) [A/cm$^2$] | Fill factor ($FF$) | Efficiency ($\eta$)[%] |
|---|---|---|---|---|
| Ag/ Cu$_2$O(493 nm thickness)/ Cu(0.5 mm thickness) | 0.26 | 1.32 | 0.47 | 0.16 |
| Ag/Cu$_2$O(493nm thickness)/ Cu(100)(5 mm thickness) | 0.71 | 9.26 | 0.42 | 2.71 |

The analysis of the values provided in Table 1 shows the increases of $\eta$ to 0.16% and 2,71%, respectively. Thus, in this case the solar cell based on the crystalline copper is optimal.

## 5. Summary

In the course of this study it has been demonstrated that the Ag/Cu$_2$O/Cu(100) solar cell based on the crystalline copper has better electrical parameters than the Ag/Cu$_2$O/Cu solar cell based on the polycrystalline copper. However, the concept of blue-shift behavior leads to a fundamentally new way of thinking about the unexpected behavior and offers a possibility of more efficient photovoltaic conversion.